\documentclass[nofootinbib,aps]{revtex4}
\usepackage{amstext,amsmath,amssymb}
\usepackage[dvips]{graphicx}
\usepackage{latexsym}

\def\be{\begin{equation}}
\def\ee{\end{equation}}
\def\bes{\begin{eqnarray}}
\def\ees{\end{eqnarray}}
\def\nn{\nonumber}

\newcommand{\SU}{\mathrm{SU}}
\newcommand{\U}{\mathrm{U}}

\def\kk{{\cal K}}
\def\pp{{\cal P}}

\def\vv{{\cal V}}
\def\dd{{\cal D}}
\def\nn{{\cal N}}
\def\sca{\mathfrak{a}}
\def\scb{\mathfrak{b}}
\def\scc{\mathfrak{c}}

\def\ga{\alpha}
\def\gd{\delta}
\def\gep{\epsilon}
\newcommand{\ft}[4]{\phi({#1}_{#2},{#1}_{#3},{#1}_{#4})}
\newcommand{\ftmon}{\phi(u_1h_\theta u_1^{-1}g_1,g_2,g_3)}
\newcommand{\fs}[5]{\phi({#1}_{#3},{#1}_{#4},{#1}_{#5};{#2}_{#3},{#2}_{#4},{#2}_{#5})}
\newcommand{\fsg}{\int_{\SU(2)} d\ga \;
\phi(g_1\alpha,g_2\alpha,g_3\alpha;u_1,u_2,u_3)}
\newcommand{\fseq}{\int_{H^3} \prod_{i=1}^{3} db_i \;
\phi(g_1,g_2,g_3;u_1b_1,u_2b_2,u_3b_3)}
\newcommand{\fsmon}{\phi(u_1h_{\theta}u_1^{-1}g_1,g_2,g_3;u_1\epsilon,u_2,u_3)}
\newcommand{\fsim}[2]{\phi({#1},{#2})}

\newcommand{\dk}[4]{\delta({#1}_{#2}{#3}_{#4}^{-1})}
\newcommand{\dkm}[6]{\delta({#1}_{#2}h_{\theta}{#1}_{#2}^{-1}{#3}_{#4}{#5}_{#6}^{-1})}
\newcommand{\dkue}[6]{\delta({#1}_{#2}{#3}_{#4}\epsilon{#5}_{#6}^{-1})}
\newcommand{\dku}[6]{\delta({#1}_{#2}{#3}_{#4}{#5}_{#6}^{-1})}
\newcommand{\dv}[8]{\delta({#1}_{#2}{#3}_{#4}^{-1}{#5}_{#6}{#7}_{#8}^{-1})}
\begin{document}
\title{A group field theory for 3d quantum gravity coupled to a scalar field}
\author{{\bf Laurent Freidel}\footnote{lfreidel@perimeterinstitute.ca}}
\affiliation{Perimeter Institute, 35 King Street North Waterloo, N2J 2W9,
Ontario, Canada.  \\
 Laboratoire de Physique, \'Ecole Normale Sup\'erieure de Lyon,
46 all\'ee d'Italie, Lyon 69007, France}
\author{{\bf Daniele Oriti}\footnote{d.oriti@damtp.cam.ac.uk}}
\author{{\bf James Ryan}\footnote{j.p.ryan@damtp.cam.ac.uk}}
\affiliation{Department of Applied Mathematics and Theoretical
Physics, Centre for Mathematical Sciences, University of
Cambridge, Wilberforce Road, Cambridge CB3 0WA, UK}
\begin{abstract}
\begin{center}
{\small ABSTRACT}
\end{center}
 We present a new group field theory model, generalising the Boulatov model,
 which incorporates both 3-dimensional gravity and matter coupled to gravity.
 We show that the Feynman diagram amplitudes of  this model are given by Riemannian quantum gravity
 spin foam amplitudes coupled to a scalar matter field.
 We briefly discuss the features of this model and its possible generalisations.
\end{abstract}
\maketitle
\section{Introduction}
Spin foam models \cite{daniele, ale} represent a purely
combinatorial and algebraic implementation of the
sum-over-histories approach to quantum gravity, in any signature
and spacetime dimension, with an abstract 2-complex playing the
role of a discrete spacetime, and algebraic data from the
representation theory of the Lorentz group playing the role of
geometric data assigned to it. Indeed, the first model of quantum
gravity to be ever proposed, the Ponzano-Regge model, was a
spin foam model for Euclidean quantum gravity without cosmological
constant \cite{PR0}. This approach has recently
been developed to a great extent in the 3-dimensional case.
 It is now clear that it provides a full quantisation of pure
gravity \cite{PR1}, whose relation with the one obtained by other
approaches is well understood \cite{PR2,lec}.
 Moreover, matter can be consistently included in the picture \cite{PR1,barrettfeynman}, providing
  a link between spin foam models and effective field theory
\cite{PR3} living on a non-commutative geometry. This picture
allows us to naturally address the semi-classical limit of spin foam
models and shows that quantum gravity in dimension $3$ effectively
follows the principle of the so-called deformed (or doubly)
special relativity \cite{dsr}.

The group field theory formalism \cite{laurentgft} represents a
generalisation of matrix models of 2-dimensional quantum gravity
\cite{mmreview}. It is a universal structure lying behind any spin
foam model for quantum gravity \cite{DP-F-K-R,carlomike}, providing a third
quantisation point of view on gravity \cite{laurentgft} and allowing
us to sum over
pure quantum gravity amplitudes associated with different topologies \cite{NPsum}. In this
picture, spin foams, and thus spacetime itself, appear as
(higher-dimensional analogues of) Feynman diagrams of a field
theory defined on a group manifold and spin foam amplitudes are
simply the Feynman amplitudes weighting the different graphs in
the perturbative expansion of the quantum field theory.
On the other hand, we can construct a non commutative field theory
whose Feynman diagram amplitudes reproduce the coupling of matter fields
 to 3d quantum gravity for a trivial topology of spacetime \cite{PR3}.
Remarkably, the momenta of the fields are labelled also by group
elements. Moreover, in three dimensions there is a duality between
matter and geometry, and the insertion of matter can be understood
as the insertion of a topological defect charged under the
Poincar\'e group \cite{PR1}. This suggests that one should be able
to treat the third quantisation of gravity and the second
quantisation of matter fields in one stroke (see \cite{mikovic}
for an early attempt). The purpose of this paper is to study how
the coupling of matter to quantum gravity is realised in the group
field theory, and whether it is possible to write down a group
field theory for gravity and particles that reproduces the
amplitudes derived in \cite{PR1} coupling quantum matter to
quantum geometry.

This is what we achieve in the present work. The way the correct
amplitudes are generated as Feynman amplitudes of the group field
theory is highly non-trivial. It requires an extension of the
usual group field theory (gft) formalism  to a higher number of
field variables, and produces an interesting intertwining of
gravity and matter degrees of freedom, as we are going to discuss
in the following. The formalism we present is still based on the
classical $SU(2)$, it bears strong similarity with a recent work
of Krasnov \cite{Kirill} who considered gft based on the quantum
group $DSU(2)$. This should not be a surprise since it is well
understood that the particle spin foam amplitudes have a quantum
group structure hidden in them
 \cite{PR1, PR2}.
The gft model we propose here is, however, very different from the
ones considered in Krasnov's work since our Feynman graphs reproduce
 explicitly the insertion of particles coupled to gravity.
\section{The Ponzano-Regge spin foam model coupled to point particles}
The general form of Feynman graph amplitudes for spinning
particles coupled to 3 dimensional quantum gravity -eg the
Ponzano-Regge model- has been written in \cite{PR1}. In this paper
we focus on the case of spinless particles and we recall in this
section the definition of these amplitudes before deriving them
from a gft.
We start from a triangulation $\Delta$ of our spacetime $M$ and
consider also the dual $\Delta^*$: dual vertices, edges and  faces
correspond  respectively to tetrahedra, faces and edges of
$\Delta$. We choose our Feynman graph, $\daleth$, to be embedded in the
triangulation $\Delta$ such that edges of $\daleth$ are edges of
the triangulation. Each edge of $\daleth$ is labelled by an angle
$\theta \in [0,\pi]$
$$ \theta= \kappa m , \qquad \kappa= 4\pi G_N \label{kappa}, $$
where $G_N$ is Newton's constant, $\kappa$ is the inverse Planck
mass and $m$ is the mass of the particle. To each angle $\theta$
we associate an element of the Cartan subgroup $H$ of $\SU(2)$
     $$ h_\theta=\left(
     \begin{array}{cc}
     e^{i\theta} & 0 \\ 0 & e^{-i\theta}
     \end{array}
     \right), $$
which corresponds to a rotation of angle $2\theta$ around a given
axis.
Given a group $G$, here $\SU(2)$, we assign group elements
$g_{e^*}$ to all dual edges $e^*$ of the triangulation. We
constrain the holonomies around dual faces $f^*\sim e$ to be flat
if there is no particle and we constrain it to be in the
conjugacy class $\theta$ if $e$ is an edge of $\daleth$. More
precisely, let us denote by $G_{e} \;(=G_{f^*})$ the product of the group
elements around a dual face (or plaquette) $f^*\sim e$:
$$G_e=G_{f^*}=\prod_{e^*\in\partial
f^*}g_{e^*}^{\epsilon_{f^*}(e^*)},$$
where $\epsilon_{f^*}(e^*)=\pm 1$ records the orientation of the
edge $e^*$ in the boundary of the (dual) face $f^*$. The amplitude
is given by
     \begin{equation} \label{PRamp1}
     {\cal Z}_M(\daleth_\theta)= \Delta(\theta)^{|E_\daleth|}
     \int\prod_{e^* }dg_{e^*}\int \prod_{e\in \daleth}du_e \,
     \prod_{e\notin \daleth}\delta(G_{e}) \prod_{e\in \daleth}
     \delta(G_{e}u_e h_\theta u_e^{-1}),
     \end{equation}
where $dg$ is the normalised Haar measure and $\delta(g)$ the
corresponding delta function on $G$,  $\Delta(\theta)\equiv
\sin(\theta)$ and $|E_{\daleth}|$ is the number of edges in the
particle graph $\daleth$. We see that two types of group elements
arise in the construction of this amplitude, the $g_{e^*}$
describe pure gravity excitations and the $u_e$ variables are
associated with the particle degrees of freedom. They arise
because the insertion of a particle locally breaks the Lorentz and
translational symmetries of the gravity model and the former gauge
transformation becomes dynamical at the location of the particle
\cite{PR1}. The $u_e$ are then interpreted as giving the direction
of the particle momenta propagating along the edge $e$. The fact
that we are talking about spinless particles implies that $u_e$
should not be thought of as an element of $G$ but as an element of
$G/H$ (with $H=U(1)$), that is momentum space. The insertion of
spinning particles can be achieved by taking into account a non
trivial dependence under the $H$ part of $u_e$. The main lesson
which we learn from this amplitude, and which gives the key idea
leading to a gft construction of such an amplitude, is the fact
that we need both $G$ variables  describing the gravity excitation
and $G/H$ variables describing the propagation of particles. We
can expand the $\delta$ functions in terms of characters
     $$
     \delta(g)=\sum_j d_j \chi_j(g),
     $$ $d_j =2j+1$
being the dimension of the spin $j$ representation and perform the
integration over $g_{e^*}$, $u_e$ in order to obtain a state sum
model
\begin{equation}\label{PRamp} {\cal Z}_M(\daleth_\theta)=\Delta(\theta)^{|E_\daleth|}
\sum_{\{j_e\}}
\prod_{e\notin \daleth} d_{j_e} \prod_{e\in
\daleth}\chi_{j_e}(h_{\theta_e})
\prod_t \left\{
\begin{array}{ccc}
    j_{e_{t_{1}}} &  j_{e_{t_{2}}} &  j_{e_{t_{3}}} \\
    j_{e_{t_{4}}} &  j_{e_{t_{5}}} &  j_{e_{t_{6}}}
    \end{array}
    \right\},
\end{equation}
where the summation is over all edges of $\Delta$ and the product
of normalised 6j symbols is over all tetrahedra $t$. For each
tetrahedron, the admissible triples of edges, e.g.
$(j_{e_{t_{1}}},j_{e_{t_{2}}},j_{e_{t_{3}}})$, corresponds to
faces of this tetrahedra. Boulatov \cite{Boul} was the first to
show that the amplitude (\ref{PRamp1}) can be obtained as a
Feynman graph evaluation of a group field theory. It is important
to note however that this amplitude is generically divergent. It
is now understood \cite{diffeo} that this divergence is due to a
translational gauge symmetry (equivalent on-shell to
diffeomorphism symmetry) acting on the Ponzano-Regge model. This
symmetry should be gauge-fixed in order to obtain well defined and
triangulation independent amplitudes. The gauge-fixing is easily
implemented by choosing a maximal tree\footnotemark
\footnotetext{A connected set of edges of $\Delta \backslash
\daleth$ passing through every vertex of $\Delta\backslash
\daleth$} $T$ of $\Delta\backslash \daleth$. The gauge-fixed
amplitude can be obtained from (\ref{PRamp1}) by replacing the
product $ \prod_{e\notin \daleth}\delta(G_{e})$ by a product over
delta functions with $e \notin T\cup \daleth$. In terms of the
state sum model (\ref{PRamp1}) the gauge-fixing inserts in the
summation a factor $\prod_{e\in T} \delta_{j_e,0}$ which
eliminates the sum over $j_e$, $e\in T$. The overall amplitude
does not depend on the choice of $T$.
\section{A gft model for 3d quantum gravity coupled to scalar
matter}\label{sec:model}
\subsection{Action and Feynman rules}
We shall now define a field theory on a group manifold, whose
Feynman expansion gives the above modified Ponzano-Regge model. We
consider a generic real field $\phi$, over the Cartesian product
of six copies of $\SU(2)$
     \begin{equation}\label{f6}
     \fs{g}{u}{1}{2}{3}:\underbrace{\SU(2)\times\dots\times
     \SU(2)}_{6}\rightarrow\mathbb{R}.
     \end{equation}
This is the basic object of the theory and, just as in the other
group field formulations of spin foam models, it has the
interpretation of a \lq 3rd quantised' chunk of quantum geometry
\cite{DP-F-K-R,laurentgft}. However, in this extended formulation
based on a 6-argument field, this chunk of quantum geometry
carries also additional degrees of freedom, labelled by the extra
$u$ variables, that acquire the physical meaning of particle
degrees of freedom (more precisely particle momenta) when a mass
parameter is inserted in a suitable way, as we are going to show
in the following. Let us now list the symmetries that this field
is required to satisfy.
\begin{itemize}
\item We require that $\phi$ is invariant under (even) elements
  $\sigma$ of the permutation group of three elements $S_3$, acting on
  {\it pairs} of field variables $(g_i,u_i)$:
     \begin{equation}\label{f6perm}
     \fs{g}{u}{1}{2}{3}=\fs{g}{u}{\sigma(1)}{\sigma(2)}{\sigma(3)}.
     \end{equation}\\
 If we require the field to be
invariant under {\it even} permutations of the three pairs of
arguments, then this is equivalent to dealing with a complex field
instead, with the odd permutations mapping the field to its
complex conjugate \cite{DP-P}, and the Feynman amplitudes produced
by the corresponding group field theory are in one-to-one
correspondence with {\it orientable} 2-complexes, as explained in
\cite{DP-F-K-R,DP-P}. We can more generally require the field to
transform under an arbitrary representation (not necessarily
reducible) of $S_3$. This will affect the type of 2-complexes
generated by the perturbative expansion of the theory. We
stress, however, that this would not imply any change for the amplitudes of
the Feynman diagrams.
\item Furthermore, we pick a $\U(1)$ subgroup $H$, of $\SU(2)$, with
  the interpretation of the
  invariance subgroup for the particle momenta, and
  project three of the arguments into $\SU(2)/\U(1)$ equivalence
  classes
     \begin{equation}\label{f6eq}
     P_b\fs{g}{u}{1}{2}{3}\equiv\fseq,
     \end{equation}
so that the field becomes in fact a function over three copies
each of $\SU(2)$ and $\SU(2)/\U(1)$.
\item Finally, we project the first half of the field, i.e. the part
  dependent on the first three arguments, into its $SU(2)$
  invariant part, by imposing invariance under simultaneous right
  action of $\SU(2)$ on the first three arguments:
     \begin{equation}\label{f6gauge}
     P_{\ga}\fs{g}{u}{1}{2}{3} \equiv \fsg.
     \end{equation}
This last symmetry has a geometric interpretation, as in the
usual Boulatov model. It imposes the closure of the triangle of
which the field $\phi$ represents the 2nd quantisation, by constraining the spin variables
dual to the $g_i$ variables associated to its three edges.
\end{itemize}
Given such a field, we can write down a Boulatov-like action, with
the extra $u$ variables simply mimicking the relations among the
gravity degrees of freedom $g$:
     \begin{equation}\label{a6pure}
     \begin{split}
     S[\phi]=
     &\frac{1}{2}\int \prod_{i=1}^{3} dg_i du_i 
     [P_{\ga}P_{b}\fs{g}{u}{1}{2}{3}][P_{\bar{\ga}}P_{\bar{b}}\fs{g}{u}{1}{2}{3}]\\
     &+\frac{\lambda}{4!}\int\prod_{i=1}^{6}dg_i du_i
[P_{\ga_1}P_{b_1}\fs{g}{u}{1}{2}{3}]
     [P_{\ga_2}P_{b_2}\fs{g}{u}{4}{5}{3}]\\
&\hphantom{xxxxxxxxxxxxxxxxxxxxx}\times[P_{\ga_3}P_{b_3}\fs{g}{u}{4}{2}{6}]
[P_{\ga_4}P_{b_4}\fs{g}{u}{1}{5}{6}].
     \end{split}
     \end{equation}
As we will see the Feynman amplitudes obtained from this model  are
proportional to those obtained by the Boulatov model, i.e. the
usual Ponzano-Regge spin foam amplitudes describing pure 3d
Riemannian quantum gravity.
 This shows that the $u$ variables are completely
redundant at this stage and do not have any real physical meaning.
They are going to acquire it soon, however.
Now we introduce a mass parameter in the theory, turning this
redundant description of pure quantum gravity into a model for
gravity coupled to scalar matter. We define a mass insertion
operator $P_\theta$, acting on the field $\phi$ as follows:
     \begin{equation}
     P_{\theta}\,\fs{g}{u}{1}{2}{3} \equiv  \;\fsmon,
     \end{equation}
where $h_{\theta}=exp(\theta J_0)\in H$; $\theta$ is half the
deficit angle created by the presence of a mass $m$, $\theta=4\pi
Gm$, $\frac{1}{4\pi G}$ being the Planck mass; $J_0$ is the
generator of the $\U(1)$ subgroup $H$, the same subgroup under
which the $u_i$ variables of the field are invariant and
$\epsilon$ is the non-trivial Weyl group element, given in the
fundamental representation by:
\begin{displaymath}
\gep=\left(\begin{array}{cc} 0 & -1\\ 1 & 0
\end{array}\right).
\end{displaymath}
We define then a new group field theory model, representing 3d
Riemannian quantum gravity coupled to scalar matter, whose
dynamics are given by the action:
     \begin{equation}\label{a6}
     \begin{split}
     S[\phi]=
     &\frac{1}{2}\int \prod_{i=1}^{3} dg_i du_i
     \bigl([P_{\ga}P_{b}\fs{g}{u}{1}{2}{3}][P_{\bar{\ga}}P_{\bar{b}}\fs{g}{u}{1}{2}{3}]\\
     &\hphantom{xxxxxxxxxxxxxxxxx}-\sca [P_{\ga}P_{b}\fs{g}{u}{1}{2}{3}][P_{\theta}P_{\bar{\ga}}P_{\bar{b}}\fs{g}{u}{1}{2}{3}]\bigr)\\
     &+\frac{\lambda}{4!}\int\prod_{i=1}^{6}dg_i du_i
     [P_{\ga_1}P_{b_1}\fs{g}{u}{1}{2}{3}]
     [P_{\ga_2}P_{b_2}\fs{g}{u}{4}{5}{3}]\\
     &\hphantom{xxxxxxxxxxxxxxxxx}\times[P_{\ga_3}P_{b_3}\fs{g}{u}{4}{2}{6}]
     [P_{\ga_4}P_{b_4}\fs{g}{u}{1}{5}{6}],
     \end{split}
     \end{equation}
where $\sca$ is a free parameter and all integrals are with
respect to the normalised Haar measure.
Let us write the Feynman rules of the theory in coordinate space.
We have to identify kinetic and vertex operators. For this
purpose, we write the action as
     \begin{equation}\label{a6op}
     \begin{split}
     S[\phi]=
     &\frac{1}{2}\int\prod_{i=1}^{3}dg_i du_i \prod_{j=1}^{3} d\bar{g}_j
     d\bar{u}_j \; \fsim{g_i}{u_i} \kk(g_i,\bar{g}_j,u_i,\bar{u}_j)
     \fsim{\bar{g}_j}{\bar{u}_j}\\
     & +\int\prod_{i,j}dg_{ij} du_{ij} \; \vv(g_{ij},u_{ij})
     \fsim{g_{1j}}{u_{2j}}\fsim{g_{2j}}{u_{2j}}\fsim{g_{3j}}{u_{3j}}\fsim{g_{4j}}{u_{4j}}.
     \end{split}
     \end{equation}
where in this integral, $i\ne j$, and
$\fsim{g_{1j}}{u_{1j}}=\fs{g}{u}{11}{12}{13}$, and so forth.
The kinetic and vertex operators are
     \begin{equation}\label{k6op}
     \kk(g_i,\bar{g}_j,u_i,\bar{u}_j) = \int\prod_{i=1}^{3}
     db_i \; \dk{g}{i}{\bar{g}}{i}
     \dku{u}{i}{b}{i}{\bar{u}}{i}
     -\sca\, \int\prod_{i=1}^{3}
     db_i \;  \dkm{u}{1}{g}{1}{\bar{g}}{1}
     \dkue{u}{1}{b}{1}{\bar{u}}{1}\prod_{j=2}^{3} \dk{g}{j}{\bar{g}}{j}
     \dku{u}{j}{b}{j}{\bar{u}}{j},
     \end{equation}
     \begin{equation}\label{v6op}
     \vv(g_{ij},u_{ij}) = \frac{\lambda}{4!}\int \prod_{i=1}^{4}d\ga_i
     \prod_{j>i} db_{ij}\;
     \dv{\ga}{j}{g}{ji}{g}{ij}{\ga}{i}
     \dv{b}{ji}{u}{ji}{u}{ij}{b}{ij}.
     \end{equation}
We have purposefully discarded the $\alpha$ variables in the
$\delta$-functions of the kinetic term. This does not change, up
to an overall factor, the computation of the amplitudes  since
$P_\theta$ commutes with $P_\alpha$. We have also sidelined the
sum over permutations, for ease of notation. Care should be taken
in inverting the kinetic term, however, on the subspace of
symmetric fields only. On this subspace, the kinetic term is
indeed diagonal and we can proceed as follows.
We now define the operators $I$ and $K_{\theta}$:
     \begin{equation}\label{k6def}
     \kk(g_i,\bar{g}_j,u_i,\bar{u}_j)\equiv
     I-\sca\;K_{\theta}.
     \end{equation}
The propagator is the inverse of the kinetic operator.
Furthermore, the operator $K_{\theta}$ satisfies
$(K_{\theta})^2=I$, as laid out below:
\begin{equation}
\begin{split}
     (K_{\theta})^2 &= \int d^3\tilde{g}\,d^3\tilde{u}\;
     K_{\theta}(g,\tilde{g},u,\tilde{u})
     K_{\theta}(\tilde{g},\bar{g},\tilde{u},\bar{u})\\
     &= \int
     d^3\tilde{g}\,d^3\tilde{u}\,d^3b\,d^3\tilde{b}\;\dkm{u}{1}{g}{1}{\tilde{g}}{1}
     \dk{g}{2}{\tilde{g}}{2}\dk{g}{3}{\tilde{g}}{3}
     \dkue{u}{1}{b}{1}{\tilde{u}}{1}\dku{u}{2}{b}{2}{\tilde{u}}{2}\dku{u}{3}{b}{3}{\tilde{u}}{3}\\
     &\hphantom{xxxxxxxxxxxxxxxxx}\times\dkm{\tilde{u}}{1}{\tilde{g}}{1}{\bar{g}}{1}
     \dk{\tilde{g}}{2}{\bar{g}}{2}\dk{\tilde{g}}{3}{\bar{g}}{3}
     \dkue{\tilde{u}}{1}{\tilde{b}}{1}{\bar{u}}{1}\dku{\tilde{u}}{2}{\tilde{b}}{2}{\bar{u}}{2}\dku{\tilde{u}}{3}{\tilde{b}}{3}{\bar{u}}{3}.
\end{split}
\end{equation}
We integrate with respect to the $\tilde{g}$ variables
\begin{equation}
\begin{split}
(K_{\theta})^2=\int d^3\tilde{u}\,d^3b\,d^3\tilde{b}\;
  \gd(\tilde{u}_1h_\theta &
\tilde{u}_1^{-1} u_1 h_\theta  u_1^{-1}g_1\bar{g}_1^{-1})
\dk{g}{2}{\bar{g}}{2}\dk{g}{3}{\bar{g}}{3}\\
&\times\dkue{u}{1}{b}{1}{\tilde{u}}{1}
\dku{u}{2}{b}{2}{\tilde{u}}{2}\dku{u}{3}{b}{3}{\tilde{u}}{3}\dkue{\tilde{u}}{1}{\tilde{b}}{1}{\bar{u}}{1}
\dku{\tilde{u}}{2}{\tilde{b}}{2}{\bar{u}}{2}\dku{\tilde{u}}{3}{\tilde{b}}{3}{\bar{u}}{3},
\end{split}
\end{equation}
and then the $\tilde{u}$ variables
\begin{equation}
\begin{split}
(K_{\theta})^2=\int d^3b\,d^3\tilde{b}\; \gd(u_1b_1\gep h_\theta
\gep^{-1} &  b_1^{-1}u_1^{-1} u_1 h_\theta u_1^{-1}g_1\bar{g}_1^{-1})
\dk{g}{2}{\bar{g}}{2}\dk{g}{3}{\bar{g}}{3}\\ &\times\gd(u_1b_1\gep
\tilde{b}_1 \gep \bar{u}_1^{-1})
\gd(u_2b_2\tilde{b}_2\bar{u}_2^{-1})\gd(u_3b_3\tilde{b}_3\bar{u}_3^{-1}).
\end{split}
\end{equation}
But $b_1\gep h_\theta \gep^{-1}b_1^{-1} =
b_1h_\theta^{-1}b_1^{-1} = h_\theta^{-1}$ since $\epsilon
h_{\theta}\epsilon^{-1}= h_\theta^{-1}$ and since  $b_1$ and $h_\theta^{-1}$ are in
the same commutative $U(1)$ subgroup. Furthermore $\gep \tilde{b}_1 \gep =
-\tilde{b}_1^{-1}$. Finally, redefining $(-b_1\tilde{b}_1^{-1})\rightarrow
b_1$, $b_2\tilde{b}_2\rightarrow b_2$ and
$b_3\tilde{b}_3\rightarrow b_3$ gives us
\begin{equation}
(K_{\theta})^2=\int d^3b\;
\dk{g}{1}{\bar{g}}{1}\dk{g}{2}{\bar{g}}{2}\dk{g}{3}{\bar{g}}{3}
\dku{u}{1}{b}{1}{\bar{u}}{1}\dku{u}{2}{b}{2}{\bar{u}}{2}\dku{u}{3}{b}{3}{\bar{u}}{3}=I.
\end{equation}
This leads to a nice closed form for the propagator
     \begin{equation}\label{p6}
     \pp(g_i,\bar{g}_j,u_i,\bar{u}_j) = 
     \frac{I+\sca K_{\theta}}{1-\sca^2}.
     \end{equation}

\subsection{Feynman amplitudes and spin foam formulation}
To construct a generic Feynman amplitude, we will analyse the
structure and gluing properties of the propagator and vertex
operator:
\subsubsection{Vertex Operator}
We scrutinise (\ref{v6op}) in two parts:
     \begin{equation}\nonumber
     \vv(g_{ij},u_{ij}) = \frac{\lambda}{4!}\int \prod_{i=1}^{4}d\ga_i
     \prod_{j>i} db_{ij}\;
     \underbrace{\dv{\ga}{j}{g}{ji}{g}{ij}{\ga}{i}}_{\text{g variable part}}
     \underbrace{\dv{b}{ji}{u}{ji}{u}{ij}{b}{ij}}_{\text{u variable part}}.
     \end{equation}
The $\gd$-functions over the $g$ variables are the usual holonomies
around the six wedges dual to the edges $e$, of a tetrahedron. So
the model already has the structure of a 2-complex dual to a
triangulation.
The $u$ variables represent the momenta of the particles and as such
are identified with the edges of the tetrahedron. Each edge of the
tetrahedron is shared by two triangles.   The $\gd$-functions
above ensure that the momentum associated to an edge is the same
when viewed from either of these triangles.  This extra
structure is not present in the Boulatov model.

\subsubsection{Propagator}
The operator (\ref{p6}) glues two tetrahedra at a triangular
interface.  From the analysis of the vertex operator above, we
know that each triangle of a tetrahedron has three wedges (dual to each of its three edges) and three
momenta associated to it. The propagator has two terms with
different gluing properties.
     \begin{equation}\nonumber
     \pp(g_i,\bar{g}_j,u_i,\bar{u}_j) =
     \underbrace{\frac{1}{1-\sca^2}I}_{\pp_{massless}}
     +\underbrace{\frac{\sca}{1-\sca^2}K_{\theta}}_{\pp_{massive}}.
     \end{equation}
For $\pp_{massless}$, the $\gd$-functions over the $g$ variables
glue three wedges of one tetrahedron to three wedges of another
tetrahedron pairwise, giving the holonomies around three composite
wedges. The $\gd$-functions over the $u$ variables ensure that the
momenta on the edges of the two triangles match.
For $\pp_{massive}$, two of the $\gd$-functions over the
$g$-variables act as for $\pp_{massless}$,  similarly for the $u$
variables.  One of the $\gd$-functions, however, couples the $g_1$
and $u_1$ variables, effectively placing a massive particle at the point where the edge of the
tetrahedron intersects the composite wedge.  The final
$\gd$-function to consider, $\dkue{u}{1}{b}{1}{\bar{u}}{1}$,
places a tag on the edge $e$, of the tetrahedron with the mass
insertion. The tag ensures that if another propagator further
around the sequence of Feynman graph edges $e^*$, forming the
boundary of a (dual) face $f^*$, inserts a mass along the edge e,
it will cancel with the first according to the property $(K_\theta)^2 = I$.
We find in general that we can
only have two possibilities when a face is fully assembled: no
particle on the edge $e$, or one particle on the edge, according
to whether there have been an even or an odd number of insertions
of $K_{\theta}$ on the boundary of the face.
To be clearer, we will calculate more explicitly the amplitude
for a generic dual face in the next subsection.
In the end the partition function for the field theory, when
expanded in Feynman graphs (i.e. in a power series for $\lambda$)
takes the form:
     \begin{equation}
     {\cal Z}=\int\dd\phi \; e^{-S[\phi]} =
     \sum_{\Gamma}\frac{\lambda^{v[\Gamma]}}{sym[\Gamma]}{\cal Z}[\Gamma],
     \end{equation}
where $v[\Gamma]$ is the number of vertices in the Feynman graph;
$sym[\Gamma]$ is its symmetry factor; and $Z[\Gamma]$ is the amplitude for each Feynman graph, being
given explicitly by:
     \begin{equation}\label{partfun}
     {\cal Z}[\Gamma]=\nn[\Gamma] \int \prod_{e^*}d\ga_{e^*} \prod_e du_e \prod_{e\notin
     \daleth}\gd(G_e)
     \prod_{e\in \daleth} \gd(G_e u_e h_\theta u_e^{-1}),
     \end{equation}
where $\nn[\Gamma]$ is a normalisation factor, discussed later, arising
from the (partial) redundancy of the $u$ variables in each
diagram and the $\sca$ dependence; $\ga_{e^*}$ is the holonomy along an edge of the Feynman
graph; $G_e =\prod_{e^*\subset \partial f^*} \alpha_{e^*}^{\pm 1}$ is
the holonomy around a face $f^*$ of the Feynman graph, which is dual to the edge of $e$ of the triangulation;
and $\daleth$ is the set of edges of the triangulation that have a
particle present.
We recognise in (\ref{partfun}) the Ponzano-Regge amplitude (\ref{PRamp1}).
We see that the group field theory we have defined gives, in
addition to a sum over all possible quantum gravity spin foams
arising as usual as Feynman graphs of the theory, a sum over all
possible massive spinless particle insertions in the spin foam,
interpreted as a sum over all possible Feynman diagrams for a
scalar field theory. Each gravity + particle configuration is
weighted exactly by the amplitude of the Ponzano-Regge model
coupled with massive spinless particles given in \cite{PR1}, and
provides us also with a definite normalisation factor for each of
these amplitudes.
\subsection{Amplitude for a generic face of the Feynman graph}
Consider a face $f^*$ of  a Feynman graph, the boundary of which
is formed by $N$ contiguous edges, $e^*$, labelled
$e_1^*,\dots,e_N^*$.
The amplitude for this face is
\begin{equation}
A(f^*)=\int(d\dots)\pp_1^{f^*}\vv_{12}^{f^*}\dots\pp_N^{f^*}\vv_{N1}^{f^*},
\end{equation}
where $\pp_i^{f^*}$ are the $\gd$-functions from the propagator
along $e_i^*$ relating to the face $f^*$; $\vv_{ij}^{f^*}$ are the
$\gd$-functions from the vertex where $e_i^*$ and $e_j^*$ meet,
pertaining to the face $f^*$.
If we contract all the $g$ variable $\gd$-functions we get a final
$\gd$-function of the form
\begin{equation}
A(f^*)= \sum_{n=0}^N \int(d\dots)\gd(G_{f^*} u_{a_1} h_\theta
u_{a_1}^{-1}\times \dots \times u_{a_n} h_\theta u_{a_n}^{-1})
\times \bigl(\text{$\gd$-functions over the $u$ variables}\bigr),
\end{equation}
 the sum is over different combinations of mass
insertions; $n\leq N$ is the number of particle insertions in that
specific term and $G_{f^*}$ is the holonomy around the face built up from
a product of $\alpha_{e^*}$.
If there are $n$ particle insertions in the $g$ variable part then
there will be $n$ $\gd$-functions in the $u$ variable part with
$\gep$ inserted.  Once we contract the $u$ variables we get that
the masses cancel just as in the calculation of $(K_\theta)^2=I$.
In the end, for each term in the sum we have two possibilities:
For $n$ even we get
\begin{equation}\label{face1}
\gd(G_{f^*})\gd(b_{f^*}),
\end{equation}
where $b_{f^*}$ is a product of the $b$ variables around the face.
Therefore we get no particle insertion and a pure gravity face
modulo an extra factor which we take into the normalisation.
For $n$ odd we get
\begin{equation}\label{face2}
\gd(G_{f^*}uh_\theta u^{-1})\gd(b_{f^*}\gep),
\end{equation}
thus a single particle insertion and another factor which we take
into the normalisation.
\subsection{Overall normalisation of the Feynman graph}
For a kinetic term with
the structure $I - \sca K_\theta$, the propagator takes the form:
$$ \frac{1}{1-\sca^2}\left( I + \sca K_\theta \right), $$ as we
have shown. This produces $\sca$-dependent amplitudes when the
expansion in Feynman graphs is performed.  We have not specified
what this $\sca$ is and how it depends on the physical parameters
of gravity or matter; indeed there is quite some freedom involved
in choosing a specific expression for $\sca$ and only further
analysis of the model we proposed can narrow the range of
possibilities down to a restricted one\footnote{a natural possibility in view of (\ref{PRamp1}) is to
consider $\sca = \Delta(\theta)$} .
This parameter will enter the normalisation coefficients controlling the relative strength
of Feynman diagrams with and without particles.
The  normalisation factor will clearly contain an overall factor
$\left(1-\sca^2\right)^{-|e^*|}$ where $|e^*|$ denotes the number of dual edges
of the two complex.
There will be an additional factor $\sca$ each time $K_\theta$ is inserted
along a dual edge.
If an even number of $K_\theta$ are inserted along a face
no particle circulates along that face.
A useful formula in order to get the right normalisation factor in a given example is
\begin{equation}
(I+\sca K_\theta )^n = \frac{1}{2}\left((1+\sca)^n+(1-\sca)^n\right) I +
\frac{1}{2}\left((1+\sca)^n-(1-\sca)^n\right)K_\theta.
\end{equation}
Along with this numerical factor, there is a singular factor coming from a redundant $\delta$-function
for each face as shown in (\ref{face1},\ref{face2}). For each face not carrying a particle we have a factor
\begin{equation}\label{fact1}
\int_{U(1)} db \,\delta(b) =\sum_{j} (2j+1),
\end{equation}
and for each face carrying a particle we have a factor
\begin{equation}\label{fact2}
\int_{U(1)} db \,\delta(b\gep) =\sum_{j} (2j+1)(-1)^j,
\end{equation}
the sum, being over integer $j$, is obtained from the character expansion of the
delta function.
These expressions are unfortunately ill defined and they need a regularisation.
A proper regularisation that can preserve all the symmetries of the theory
is to replace the usual $\SU(2)$ group by a quantum group $U_q(\SU(2))$.
It is expected that with this choice the key features of the model can be preserved and
that the  corresponding normalisation coefficients are given by
\begin{equation}
\sum_{j=0}^N [2j+1]_q t^j = \frac{1+ t + (q+q^{-1})t^{N+1}}{(1-q^{-2} t ) (1-q^2 t)},
\end{equation}
with $q =\exp(i\frac{\pi}{2N+1})$, $[n]_q = (q^n-q^{-n})/(q-q^{-1})$ and $t =-1$ for a face with
particle and $t=+1$ otherwise.
Let us recall that if we consider the original Boulatov model such
factors do not arise since they come from a redundant summation over
spins dual to the $u$ variables. The Feynman graph amplitudes of the
Boulatov model are, however, not equal to the physical quantum gravity
amplitudes. In order to get the quantum gravity amplitudes one has to
divide out the infinite volume of a gauge symmetry. This is
conveniently done by restricting the summation over spins as presented
in the first section.
This gauge-fixing procedure
which is well defined at the level of the quantum gravity amplitudes is,
however, not fully understood  at the level of the gft.
When we extend the gft to include the momenta variables $u$, we
also extend in a trivial way  the gauge symmetry of the quantum gravity
amplitude. This is where the additional factors
(\ref{fact1}, \ref{fact2}) come from.
The gauge-fixing is trivially realised by fixing to $0$  the
spin dual to the variables $u_{e}$.
This is, we feel, one of the main open challenges in this domain: to understand if such a
gauge fixing can be understood already at the gft level both for the
Boulatov model and for our extension including particles;
that is, whether we can already for the Boulatov model identify at the
level of the gft the translational (or diffeomorphism) symmetry responsible
for the divergences of the na\"ive gravity amplitude. A similar
identification should also be implemented for our particle model.
We do not resolve this issue in the present work.
\section{Discussion}
\subsection{Features of the model}
We have seen that the model correctly generates spin foam
configurations with some dual faces carrying particle data, i.e. a
mass label, indicating that a particle of the given mass is
propagating along the edge of the triangulation dual to that face.
This means that the group field theory produces all possible
Feynman graphs for a scalar field embedded in the triangulation
dual to the quantum gravity 2-complex, specifying the field
propagator on each line of the Feynman graph, and this only.
Interestingly, this is enough, in this 3-dimensional setting, for
specifying fully the dynamics of the particles, i.e. their
interaction. In fact, this is dictated by the Bianchi identity
constraining the sum of curvatures in the boundary of any 3-cell
of the dual complex around any given vertex of the triangulation.
When one or more particles are meeting at that vertex, thus
interacting there, this implies momentum conservation for their
interaction, which is the only content of any $\phi^n$ theory.
Because any number of particles can be incident to any given
vertex of the triangulation in the model we propose, this means
that this corresponds to a scalar field theory with a potential
given by a sum over any power of the field operators: $\phi^3(x) +
\phi^4(x) +....$.

We have seen that the crucial property of the modified kinetic
term we propose for producing mass insertions in the spin foam
amplitudes is, besides the extension of the field to 6 arguments,
the property $(K_\theta)^2=I$ for the operator $K_\theta$ inserting
the mass of the particles in the group field theory action.
It is interesting to note that the presence of this operator which
inserts particles also breaks a symmetry that the pure gravity model possesses.
This bears some similarity with the fact that particles in 3d can be
understood as defects breaking the translational symmetry of the theory
without matter \cite{PR1}.
The symmetry is the following:  Let's consider the transformation
\begin{equation}
\phi(g_1,g_2,g_3;u_1,u_2,u_3)
\rightarrow
\phi({\mathbf v_1}g_1,g_2,g_3;{\mathbf w_1}u_1,u_2,u_3),
\end{equation}
where ${\mathbf v_1},{\mathbf w_1}$ are arbitrary fixed group elements.
This transformation is clearly a symmetry of the pure gravity action
(\ref{a6pure}).
This symmetry is, however, broken by the insertion of a mass term and only the
transformation
\begin{equation}
\phi(g_1,g_2,g_3;u_1,u_2,u_3)
\rightarrow
\phi({\mathbf v_1}g_1,g_2,g_3;{\mathbf v_1}u_1,u_2,u_3),
\end{equation}
preserves the action (\ref{a6}).
\subsection{A direct generalisation}
The model we presented in section \ref{sec:model} produces mass
insertions in different faces of the spin foam 2-complex by
application of the operator $K_\theta$ in the 1st argument of the
field, carrying the gravity variable $g_1$. Because of permutation
symmetry, the fact that one has chosen the 1st argument of the
field for inserting a mass parameter and not, say, the 2nd is
irrelevant, as one can easily convince oneself. Still, one may
find the fact that a mass parameter is inserted in only -one- of
the arguments of the field a bit unsatisfactory, for symmetry
reasons. Here we want to discuss briefly what happens when one
relaxes this condition. The result is that one can write down a
generalised version of the model presented above, that is, however,
basically equivalent to it, and leads to the same type of graphs
being generated.
One can consider defining a generalised kinetic term with matter
insertions, defined using a sum of operators
$K_\theta(1)$, $K_\theta(2)$, $K_\theta(3)$, each $K_\theta(i)$
inserting a mass parameter in the i-th argument of field, thus
having as kinetic term an operator with the structure $I
-\sca(K_\theta(1) + K_\theta(2) +K_\theta(3))$. It is obvious that
a model like this would generate exactly the same type of graphs
and amplitudes as the one we have defined above. It is also easy
to realise that, once such an operator is introduced, there will
be graphs in which the insertion of a $K_\theta(1)$ and a
$K_\theta(2)$, say, in different propagators would produce the
same amplitude that would have been generated by the use in the
kinetic term of an operator of the form $K_\theta(1,2)$, i.e. an
operator inserting a mass parameter in both the 1st and 2nd
arguments of the field at once; and indeed one could generalise
further the kinetic term to an operator of the form: $I - \sca (
K_\theta(1) + K_\theta(2) +K_\theta(3)) - \scb (  K_\theta(1,2) +
K_\theta(2,3) +K_\theta(1,3))$. Carrying this line of reasoning
even further, one is led to the kinetic term:
     \begin{equation}\label{genk}
     \kk = I- \sca(K_\theta(1) + K_\theta(2) + K_\theta(3)) -
     \scb(K_\theta(1,2) + K_\theta(2,3) + K_\theta(1,3)) -
     \scc K_\theta(1,2,3) \equiv I-K_{total},
     \end{equation}
where the $K_\theta(1,2,3)$ is defined as the operator
inserting a mass parameter in all the first three arguments of the
field. Again, this generalised kinetic term leads to the same kind of Feynman graphs
 and amplitudes, as it is easy to verify.
 In fact the structure of the amplitudes is determined by the property $K_\theta(i)^2=1$,
 as we have explained, so it is enough for each mass insertion produced by the
operators $K_\theta(i,j)$ and $K_\theta(i,j,k)$ to satisfy that
property in order for the resulting amplitude to be of the form we have
described. Of course, the normalisation factors for the amplitudes
are going to be different from those of the simpler model
presented in section \ref{sec:model} and it will depend in general
on three {\it different} coupling constants $\sca$,
$\scb$ and $\scc$. This gives additional freedom that may well turn out
to be useful in some situation.

It is interesting to note that there are several choices of coupling constant that lead to further
simplifications.
For example, one can show that in order for the added terms to satisfy $(K_{total})^2 \propto I$, then one needs to
choose
     \begin{equation}\label{con}
     \sca=0,\quad \scb=0,\quad \mathrm{or}\quad  \scb=0,\quad \sca+\scc=0,
     \end{equation}
that is
     \begin{equation}\label{kin}
     \kk  =  I - \scc K_\theta(1,2,3),\quad \mathrm{or}\quad  \kk  =  I - \sca\left(
     K_\theta(1) + K_\theta(2) + K_\theta(3)- K_\theta(1,2,3)\right).
     \end{equation}
Finally, a  simple and highly symmetrical choice is
     \begin{equation}\label{symk}
\begin{split}
     \kk &= (I - \sca K_\theta(1))(I - \sca K_\theta(2))(I - \sca K_\theta(3))\\
&= I- \sca(K_\theta(1) + K_\theta(2) + K_\theta(3)) +
     \sca^2(K_\theta(1,2) + K_\theta(2,3) + K_\theta(1,3)) -
     \sca^3K_\theta(1,2,3).
\end{split}
     \end{equation}
Since the $(I - \sca K_\theta(i))$  commute, we can easily compute
the propagator
\begin{equation}
\pp = \frac{(I + \sca K_\theta(1))(I + \sca K_\theta(2))(I + \sca K_\theta(3))}{(1-\sca^2)^3}.
\end{equation}
It is not clear at the present stage, however, which specific properties one should ask the gft propagator to
fulfill.
\subsection{Alternatives}
After having discussed some possible generalisations of the
proposed model leading to very similar structures, we would like
to discuss two genuine alternatives to it. One based on a much
simpler action constructed inserting a mass parameter in the
simplest way in the usual 3-argument field, leading however to a
problematic structure for the Feynman amplitudes, and thus showing
the need for the 6-argument extension on which we have based our model.
The other keeps the same structure of the model presented in
section \ref{sec:model}, but inserts the mass parameter by means
of a modification of the interaction term in the gft action,
instead of the kinetic one. As we will show, this modification is
completely harmless.
Consider first a three argument field, the same on which the
Boulatov model is based, and insert the mass parameter $h_{\theta}$ and the
velocities for the relevant particles $u_i$ in the 1st argument of
the field. The model we obtain is therefore realised by
forgetting about the presence of the $u_i$ variables in the extra
slots of the generalised field used in the model presented in
section \ref{sec:model}. The action is then:
\begin{equation}\label{a3}
\begin{split}
S[\phi]=
\frac{1}{2}\int \prod_{i=1}^{3}dg_i du_1\;
\bigl(\ft{g}{1}{2}{3}&\ft{g}{1}{2}{3} -\sca
\;\ft{g}{1}{2}{3}\ftmon\bigr)\\
&+\frac{\lambda}{4!}\int \prod_{i=1}^{6}dg_i \;
\ft{g}{1}{2}{3}\ft{g}{4}{5}{3}\ft{g}{4}{2}{6}\ft{g}{1}{5}{6}.
\end{split}
\end{equation}
The Feynman rules for this action can be read out easily, and the
amplitudes constructed in the usual way. The kinetic operator is:
     \begin{equation}\label{k3op}
     \kk(g_i,\bar{g}_j) = \prod_{i=1}^{3}
\dk{g}{i}{\bar{g}}{i}
     -\sca \, \int du_1\;\dkm{u}{1}{g}{1}{\bar{g}}{1}
\prod_{i=2}^{3} \dk{g}{i}{\bar{g}}{i}
     \equiv I-\sca K_\theta,
     \end{equation}
while the interaction operator is the usual Boulatov one. It is
easy to see from the expression for the kinetic operator, that the
mass-inserting operator does not satisfy any property like
$(K_\theta)^2=I$ anymore. Instead, its square gives:
\begin{equation}
(K_\theta)^2= \int du_1\,d\tilde{u}_1 \gd(\tilde{u}_1 h_\theta
     \tilde{u}_1^{-1}u_1 h_\theta u_1^{-1} g_1\bar{g}_1^{-1} )
\prod_{i=2}^{3} \dk{g}{i}{\bar{g}}{i}.
\end{equation}
This makes the propagator much more complicated, being
given by:
\begin{equation}
\pp(g_i,\bar{g}_j)= I +\sum_{n> 0}(\sca K_\theta)^n,
\end{equation}
resulting in considerably more arduous Feynman diagrammatics. An
example of an amplitude of this model, for a typical dual face
is:
\begin{equation}
A(f^*)= \gd(G_{f^*} u_{a_1} h_\theta u_{a_1}^{-1}\times \dots
\times u_{a_n} h_\theta u_{a_n}^{-1}),
\end{equation}
for $n$ less than the number of edges bounding the dual face
$f^*$.
We see that  there is no \lq multiple mass cancellation'
anymore, and we end up having multiple mass insertions on each
face, in the typical Feynman diagram, each of which is associated
to a different $SU(2)/U(1)$ velocity element. There is a possible
physical interpretation of the resulting configuration, which is
in terms of multi-particle states. In other words, we would have
more than one particle with a given mass associated to a dual face
and so to an edge of the triangulation. We are not in the position
of being able to exclude  this interpretation, or definitely
reject the model that generates these configurations altogether;
however, we feel that such an interpretation is problematic for at
least two reasons: 1) it would imply that more than one particle
is propagating along the same link of the triangulation, so they
would be located at the same point in the manifold and interact
together with other (multi-)particles at the vertices of the
triangulation; most important, 2) interpreting the particle
configurations in the triangulation as Feynman graphs of some
effective field theory would become much less straightforward than
for the model proposed in section \ref{sec:model}, and applying
any procedure to extract this effective field theory, like that
used in \cite{PR3}, to the amplitudes generated by this \lq
simplified' model would be quite cumbersome, if at all possible.
We note that the same problem of \lq multiple mass insertions' is
generated by most other modifications of the structure of the
model presented in section \ref{sec:model}, affecting the 3 extra
arguments of the field, i.e. the $u_i$ variables.  Such
modifications lead to losing the property $(K_\theta)^2=I$ which is
responsible for \lq mass cancellation' in the dual faces when the
operator is inserted more than once, and that leads to the
presence of only one mass in each dual face, and furthermore, to consistency with the
interpretation of the mass-labelled graphs in the triangulation as
Feynman graphs of a field theory.

We conclude by mentioning instead a harmless modification of the
model, that may even turn out to be useful in
future studies. One can keep the structure of the field to be a
function of 6 arguments, and one can keep the same form for the pure
gravity action based on this 6 argument field, but choose to
insert a mass parameter not in the kinetic term but in the
interaction term of the group field theory action. The group field
theory action would then be:
     \begin{equation}
     \begin{split}
     S[\phi]=
     &\frac{1}{2}\int \prod_{i=1}^{3} dg_i\,du_i
     [P_{\ga}P_{b}\fs{g}{u}{1}{2}{3}][P_{\bar{\ga}}P_{\bar{b}}\fs{g}{u}{1}{2}{3}]\\
     &+\frac{\lambda}{4!}\int\prod_{i=1}^{6}dg_i\,du_i 
[P_{\ga_1}P_{b_1}\fs{g}{u}{1}{2}{3}]
     [P_{\ga_2}P_{b_2}\fs{g}{u}{4}{5}{3}]\\
     &\hphantom{xxxxxxxxxxxxxxxxxxxxx}\times[P_{\ga_3}P_{b_3}\fs{g}{u}{4}{2}{6}]
     [P_{\ga_4}P_{b_4}\fs{g}{u}{1}{5}{6}]\\
     &+\frac{\lambda}{4!}\int\prod_{i=1}^{6}dg_i\,du_i\;\sca
[P_{\theta}P_{\ga_1}P_{b_1}\fs{g}{u}{1}{2}{3}]
     [P_{\ga_2}P_{b_2}\fs{g}{u}{4}{5}{3}]\\
     &\hphantom{xxxxxxxxxxxxxxxxxxxxx}\times[P_{\ga_3}P_{b_3}\fs{g}{u}{4}{2}{6}]
     [P_{\ga_4}P_{b_4}\fs{g}{u}{1}{5}{6}].
     \end{split}
     \end{equation}
As we have anticipated, however, it is straightforward to verify
that this action leads to the same type of Feynman graphs and
amplitudes that one gets instead by modifying the kinetic term,
the only difference being that one would get a different
normalisation factor, in front of each amplitude.

\section{Conclusions}
We have defined a new group field theory for 3-dimensional
Riemannian quantum gravity, and constructed its perturbative
expansion in Feynman diagrams. These diagrams correspond to spin
foam 2-complexes describing quantum gravitational degrees of
freedom dual to 3d triangulations, as in the Boulatov model,
but they also carry additional labels which describe massive spinless
particles propagating in the spacetime one reconstructs from the
gravity degrees of freedom.
They have  the interpretation of
Feynman graphs for a scalar field theory embedded in the
triangulation representing spacetime. The amplitudes for these
diagrams have exactly the form obtained in \cite{PR1} from
classical considerations, i.e. are given by the Ponzano-Regge model
coupled to massive spinless particles, and are shown in \cite{PR3} to
admit an effective (non-commutative) scalar field theory
description, confirming the above physical interpretation.

The model presented possesses some quite non-trivial and interesting
features, that we highlighted in the paper, and that deserve
further analysis.
This represents a first step in a program of analysing the
coupling of matter and gauge fields to quantum gravity in the
group field theory approach. The next steps would be first of all
the inclusion of spin degrees of freedom and the construction of a
group field theory reproducing the amplitudes given in \cite{PR1}
for massive spinning particles. Second, one should develop the study of the
gft observables in the presence of particles and its relation with spin networks with open ends.
Next, one would like to show how
the effective field theory picture for the particle degrees of
freedom can be obtained directly, and possibly in a simpler way,
from the group field theory formulation. Finally, the problem of the
coupling of gauge fields and the description of their interaction
with both gravity and matter fields, again in the group field
theory formalism, should be tackled.

To conclude, as we mention in the text, one of the most pressing issues for our model and the Boulatov model is
 to understand whether there are symmetries at the gft level that justify the gauge fixing
needed at the level of the Feynman graphs to reproduce physical 3d gravity amplitudes.
This is necessary in order to promote this type of gft to a fundamental model of three dimensional gravity
coupled to matter.

{\bf Acknowledgments:} We would like to thank K. Krasnov for keeping us informed of his progress and PI
for an invitation which initiated this collaboration. J.R. would like to
thank the occupants of room B0.10 for correcting innumerable typos.

\end{document}